\documentclass[12pt,a4paper]{article}

\newcommand{\la}{\lambda}
\newcommand{\pa}{\partial}
\newcommand{\al}{\alpha}
\newcommand{\ep}{\epsilon}

\begin{document}

\begin{flushright}
{}
\end{flushright}
\vspace{1.8cm}

\begin{center}
 \textbf{\Large Conformal SO(2,4) Transformations \\
of the One-Cusp Wilson Loop Surface }
\end{center}
\vspace{1.6cm}
\begin{center}
 Shijong Ryang
\end{center}

\begin{center}
\textit{Department of Physics \\ Kyoto Prefectural University of Medicine
\\ Taishogun, Kyoto 603-8334 Japan}  \par
\texttt{ryang@koto.kpu-m.ac.jp}
\end{center}
\vspace{2.8cm}
\begin{abstract}
By applying the conformal SO(2,4) transformations to the elementary 
one-cusp Wilson loop surface we construct various two-cusp and
four-cusp Wilson loop surface configurations in $AdS_5$ and demonstrate
that they solve the string equations of the Nambu-Goto string action.
The conformal boosts of the basic four-cusp Wilson loop surface with a
square-form projection generate various four-cusp Wilson loop surfaces 
with projections of the rescaled square, the rhombus and the
trapezium, on which surfaces the classical Euclidean Nambu-Goto string
actions in the IR dimensional regularization are evaluated.
\end{abstract} 
\vspace{3cm}
\begin{flushleft}
October, 2007
\end{flushleft}

\newpage
\section{Introduction}

The AdS/CFT correspondence \cite{MGW} has more and more revealed the deep
relations between the $\mathcal{N}=4$ super Yang-Mills (SYM) theory
and the string theory in $AdS_5\times S^5$, where classical string 
solutions play an important role \cite{GKP,AT,JP}. The energies of 
classical strings have been shown to match with the anomalous dimensions
of the gauge invariant operators, while an open string ending on a curve
at the boundary of $AdS_5$ has been analyzed to study the strong coupling
behavior of the Wilson loop in the gauge theory \cite{RY,JM,DGO}.

Alday and Maldacena have used the AdS/CFT correspondence to compute the
planar 4-gluon scattering amplitude at strong coupling in the 
$\mathcal{N}=4$ SYM theory \cite{AM} and found agreement with the result
of a conjectured form regarding the all-loop iterative structure and
the IR divergence of the perturbative gluon amplitude \cite{ABD}.
The 4-gluon scattering amplitude has been evaluated as the string theory
computation of the 4-cusp Wilson loop composed of 4 lightlike segments
in the T-dual coordinates, where a certain open string solution in
$AdS_5$ space is found to minimize the area of the string surface whose
boundary conditions are determined by the massless gluon momenta, and a 
dimensional regularization is used to regularize the IR divergence.
Before the IR regularization the worldsheet surface of this particular
solution \cite{AM} is related by a certain conformal SO(2,4) 
transformation to the 1-cusp Wilson loop surface found in 
\cite{MK} (see also \cite{YM}).

The non-leading prefactor of the gluon amplitude has been studied 
\cite{AFK} and the IR structure of $n$-gluon amplitudes has been fully
extracted from a local consideration near each cusp \cite{EB},
where the 1-cusp Wilson loop solution is constructed even in the presence
of the IR regularization. By computing the 1-loop string correction to 
the 1-cusp Wilson loop solution, the 1-loop coefficient in the cusp
anomaly function $f(\la)$ of the gauge coupling $\la$ has been 
derived as consistent with the energy of a closed string
with large spin $S$ in $AdS_5$ \cite{KRT}. Moreover, the 2-loop 
coefficient in $f(\la)$ has been presented \cite{RT} to agree with
the results of \cite{BBK,BKK} for the strong coupling solution of the 
BES equation \cite{BES} in the gauge theory side. Based on the string
sigma-model action a whole class of string solution for the 4-gluon
amplitude has been constructed \cite{MMT} under the constraint that the
Lagrangian evaluated on the string solution takes a constant value.
Applying the dressing method \cite{SV} used for the study of the giant
magnons and their bound or scattering states \cite{HM,ND} to the
elementary 1-cusp Wilson loop solution of 
\cite{MK}, new classical solutions
for Euclidean worldsheets in $AdS_5$ \cite{JKS} have been constructed,
where the surfaces end on complicated, timelike curves at the boundary
of $AdS_5$. Several investigations associated with planar gluon
amplitudes have been presented \cite{BHT,DKS,KR,LAM,DHK,NSV,JMS}.

In ref. \cite{AM} the planar 4-gluon amplitude at strong coupling has been
constructed by deriving the classical string sigma-model action evaluated
on the 4-cusp Wilson loop surface whose edge traces out a rhombus on
the projected two-dimensional plane at the boundary of $AdS_5$. 
Based on the Nambu-Goto string action we will apply various conformal
SO(2,4) transformations to the elementary 1-cusp Wilson loop solution of
\cite{MK}. We will show how the obtained string surface 
configurations satisfy the string equations of motion derived from
the Nambu-Goto string action. We will observe that there appear various
kinds of Wilson loop solutions which are separated into the
2-cusp Wilson loop solutions and the 4-cusp ones.

\section{SO(2)$\times$SO(4) transformations of the 1-cusp Wilson loop
solution} 

We consider the 1-cusp Wilson loop solution of \cite{MK},
where the open string world surface ends on two semi infinite lightlike
lines and is given by
\begin{equation}
r = \sqrt{2}\sqrt{y_0^2 - y_1^2}
\label{els}\end{equation}
embedded in an $AdS_3$ subspace of $AdS_5$ with the metric written in the
T-dual coordinates  by \cite{AM}
\begin{equation}
ds^2 = \frac{-dy_0^2 + dy_1^2 + dr^2}{r^2}.
\end{equation}
Here we take the static gauge where $(y_0,y_1)$ are regarded as worldsheet
directions to write the Nambu-Goto action 
\begin{equation}
S = \frac{R^2}{2\pi}\int dy_0dy_1 \frac{1}{r^2} \sqrt{D}, \hspace{1cm}
D = 1 + \left(\frac{\pa r}{\pa y_1}\right)^2 - \left(
\frac{\pa r}{\pa y_0}\right)^2,
\end{equation}
from which the equation of motion for $r$ is given by
\begin{equation}
\frac{2\sqrt{D}}{r^3} = \pa_0\left(\frac{\pa_0r}{r^2\sqrt{D}}\right) -
  \pa_1\left(\frac{\pa_1r}{r^2\sqrt{D}}\right).
\label{ste}\end{equation}
The solution (\ref{els}) is confirmed to satisfy the eq. (\ref{ste}) with
$\sqrt{D}=i$, which implies that the Lagrangian is purely imaginary when
it is evaluated on the solution (\ref{els}). Then the amplitude 
$\mathcal{A}\sim e^{iS}$ has an exponential suppression factor. The 
Poincare coordinates in $AdS_5$ with the boundary $r=0$,
\begin{equation}
ds^2 = \frac{-dy_0^2 + dy_1^2 + dy_2^2 + dy_3^2 + dr^2}{r^2}
\end{equation}
are related to the embedding coordinates $Y_M \; (M= -1,0,\cdots,4)$
on which the conformal SO(2,4) transformation is acting linearly by the
following relations 
\begin{eqnarray}
Y^{\mu} = \frac{y^{\mu}}{r}, \hspace{1cm} \mu = 0, \cdots, 3, \nonumber \\
Y_{-1} + Y_4 = \frac{1}{r}, \hspace{1cm} Y_{-1} - Y_4 = 
\frac{r^2 + y_{\mu}y^{\mu}  }{r}, \\
-Y_{-1}^2 - Y_0^2 + Y_1^2 + Y_2^2 + Y_3^2 + Y_4^2 = -1. \nonumber
\end{eqnarray}
The elementary 1-cusp solution (\ref{els}) is expressed in terms of 
$Y_M$ as
\begin{equation}
 Y_0^2 - Y_{-1}^2 = Y_1^2 - Y_4^2, \hspace{1cm} Y_2 = Y_3 = 0.
\label{emy}\end{equation}

Let us make an SO(2,4) transformation defined by 
\begin{equation}
\left(\begin{array}{c}Y_0 \\ Y_{-1} \end{array} \right) = P
\left(\begin{array}{c}Y_0' \\ Y_{-1}' \end{array} \right),
\hspace{1cm} \left(\begin{array}{c}Y_1 \\ Y_2 \\ Y_3 \\ Y_4 \end{array}
 \right) = Q \left(\begin{array}{c}Y_1' \\ Y_2' \\ Y_3' \\ Y_4' 
\end{array}  \right)
\end{equation}
with
\begin{equation}
P = \frac{1}{\sqrt{2}} \left( \begin{array}{cc} 1 & -1 \\ 1 & 1 
\end{array} \right) \equiv P_1, \hspace{1cm}
Q = \left( \begin{array}{cccc} \frac{1}{\sqrt{2}} & 0 & 0 &  
-\frac{1}{\sqrt{2}} \\ 0 & 1 & 0 & 0 \\ 0 & 0 & 1 & 0 \\
\frac{1}{\sqrt{2}} & 0 & 0 & \frac{1}{\sqrt{2}} \end{array} \right)
\equiv Q_1.
\end{equation}
This SO(2)$\times$SO(4) rotation of the elementary 1-cusp solution 
(\ref{emy}) generates a configuration 
\begin{equation}
Y_0' Y_{-1}' = Y_1' Y_4',
\end{equation}
which is equivalently expressed in terms of the Poincare coordinates as
\begin{equation}
r = \sqrt{y_0^2 - y_1^2 -  \frac{y_0 - y_1}{y_0 + y_1}  }
\equiv \sqrt{A},
\label{yrs}\end{equation}
where the prime has been suppressed for convenience. 
When the string configuration (\ref{yrs}) is substituted into
(\ref{ste}), $\sqrt{D}$ is so compactly given by $i/|y_0 + y_1|$ that
the right hand side (RHS) of (\ref{ste}) is separated into two parts 
for the region $y_0 + y_1>0$
\begin{eqnarray}
\frac{1}{iA^{3/2}} \left[ \left(y_1 + 2y_0  + \frac{y_1}{(y_0+y_1)^2} 
\right)  + \left( y_0 + 2y_1 + \frac{y_0}{(y_0+y_1)^2} \right)\right]
\nonumber \\
+ \frac{3}{2iA^{5/2}} \left[ \left( -y_0(y_0+y_1) + \frac{y_1}{y_0+y_1}
\right) \pa_0A + \left( -y_1(y_0+y_1) + \frac{y_0}{y_0+y_1}
\right) \pa_1A  \right],   
\end{eqnarray}
whose second $1/A^{5/2}$ part becomes proportional to $1/A^{3/2}$ and then
the equation of motion (\ref{ste}) is satisfied.
For the region $y_0 + y_1<0$ the string equation is similarly satisfied.
The solution (\ref{yrs}) shows that the surface ends on the lines 
specified by $y_0 = y_1, y_0 = -y_1 \pm 1$ where two cusps are located at
$(y_0,y_1) = (\pm 1/2, \pm 1/2)$.

The SO(2,4) transformations given by $P=1_2, 2\times 2$ unit matrix,
$Q =Q_1$ and $P = -i\sigma_2$ that interchanges $Y_0$ and $Y_{-1}$,
$Q=Q_1$ produce the following configurations
\begin{equation}
{Y_0'}^2 - {Y_{-1}'}^2 = - 2Y_1' Y_4', \hspace{1cm}
{Y_0'}^2 - {Y_{-1}'}^2 =  2Y_1' Y_4'
\end{equation}
respectively, which turn out to be
\begin{eqnarray}
r^2 &=& y_0^2 - ( y_1 + 1 )^2 \pm 2\sqrt{y_0^2 + (y_1 + 1)^2 - 1},
\label{rty} \\
r^2 &=& y_0^2 - ( y_1 - 1 )^2 \pm 2\sqrt{y_0^2 + (y_1 - 1)^2 - 1}.
\label{try}\end{eqnarray}
In order to show that the latter surface equation obeys the string
equation (\ref{ste}) we parametrize $r$ as $r=\sqrt{y_0^2 - (y_1 -1)^2 
+ 2\sqrt{B} } \equiv \sqrt{A}$ for the plus sign, and 
$B \equiv y_0^2 + (y_1 - 1)^2 - 1$. In this case $\sqrt{D}$ is again a 
pure imaginary $\sqrt{D} = i/\sqrt{B}$. The RHS of (\ref{ste}) has 
also two parts 
\begin{equation}
\frac{1}{iA^{3/2}\sqrt{B}}[2B + (y_1 -1)^2 + y_0^2 ]
+ \frac{3}{iA^{5/2}\sqrt{B}}[(y_1 -1)^2(1 - \sqrt{B})^2 -
 y_0^2(1 + \sqrt{B})^2 ],
\end{equation}
whose second part again becomes proportional to $1/A^{3/2}$ and combines
with the first part to yield $2i/A^{3/2}\sqrt{B}$ which is just
the left hand side (LHS) of (\ref{ste}).

For the plus sign of (\ref{try}) at the boundary of $AdS_3, r = 0$, 
the surface ends on two lines $y_0 = -y_1 + \sqrt{2} + 1, 
y_0 = y_1 - (\sqrt{2} + 1)$ in $y_1 \ge 1+ 1/\sqrt{2}$ and  
two lines $y_0 = -y_1 - (\sqrt{2} - 1), 
y_0 = y_1 + \sqrt{2} - 1$ in $y_1 \le 1 - 1/\sqrt{2}$. 
The region in the outside of the circle defined by $y_0^2 + (y_1 -1)^2
= 1$ is allowed and there are two cusps located at $(y_0,y_1) =
(0,\sqrt{2} +1), (0,-\sqrt{2} +1)$, where one semi infinite lightlike
line and one finite lightlike line meet at each cusp for $y_0 \ge 0$
and the allowed region specified by $r^2 \ge 0$ is separated into
$y_0 \ge 0$ part and $y_0 \le 0$ part. For the minus sign of (\ref{try})
the surface ends on two lines $y_0 = -y_1 + \sqrt{2} + 1, 
y_0 = y_1 + \sqrt{2} - 1$ in $y_0 \ge 1/\sqrt{2}$ and  two lines 
$y_0 = -y_1 - (\sqrt{2} - 1), y_0 = y_1 - (\sqrt{2} + 1)$ in 
$y_0 \le -1/\sqrt{2}$. There are two cusps located at $(y_0,y_1) =
(\pm\sqrt{2}, 1)$. The string surface ends on the two semi infinite 
lightlike lines which emerge from the one cusp $(\sqrt{2},1)$
for the region $y_0 \ge \sqrt{2}$. The former surface (\ref{rty})
is similarly shown to be a two-cusp Wilson loop solution. 

If the other conformal transformations are performed by $P = P_1, Q = 
1_4, 4\times 4$ unit matrix and $P = P_1$, 
\begin{equation}
Q = \left( \begin{array}{cccc} 0 & 0 & 0 & -1 \\ 0 & 1 & 0 & 0 \\
 0 & 0 & 1 & 0 \\ 1 & 0 & 0 & 0 \end{array} \right) \equiv Q_2,  
\end{equation}
that interchanges $Y_1$ and $Y_4$, we have two curves
\begin{equation}
-2Y_0' Y_{-1}' = {Y_1'}^2 - {Y_4'}^2, \hspace{1cm} 
 2Y_0' Y_{-1}' = {Y_1'}^2 - {Y_4'}^2,
\end{equation}
which are expressed in terms of the Poincare coordinates as
\begin{eqnarray}
r^2 &=& (y_0 + 1)^2 - y_1^2 \pm 2\sqrt{(y_0 + 1)^2 + y_1^2 - 1},
\label{rhy} \\
r^2 &=& (y_0 - 1)^2 - y_1^2 \pm 2\sqrt{(y_0 - 1)^2 + y_1^2 - 1},
\label{hry}\end{eqnarray}
respectively. These expressions are compared with (\ref{rty}) and 
(\ref{try}) under the exchange of $y_0$ and $y_1$.
The two curves (\ref{rhy}) and (\ref{hry}) also obey the
string eq. (\ref{ste}) in the same way as (\ref{try}).

For the plus sign of (\ref{rhy}) the surface ends on two lines 
$y_0 = -y_1 + \sqrt{2} - 1, y_0 = y_1 - (\sqrt{2} + 1)$ in $y_1 \ge 
1/\sqrt{2}$ and  two lines $y_0 = -y_1 - (\sqrt{2} + 1), y_0 = y_1 + 
\sqrt{2} - 1$ in $y_1 \le -1/\sqrt{2}$ which meet at two cusps 
$(-1,\pm\sqrt{2})$ respectively. 
For the minus sign of (\ref{rhy}) the surface ends on two lines 
$y_0 = -y_1 + \sqrt{2} - 1, y_0 = y_1 + \sqrt{2} - 1$ in $y_0 \ge 
-1 +1/\sqrt{2}$ and  two lines $y_0 = -y_1 - (\sqrt{2} + 1), y_0 = y_1 -
(\sqrt{2} + 1)$ in $y_0 \le -1 -1/\sqrt{2}$ which meet at two cusps 
$(\sqrt{2}-1,0)$ and  $(-\sqrt{2}-1,0)$ respectively. 
Similarly for the plus sign of (\ref{hry}) two cusps are located at
$(1, \pm\sqrt{2})$ and for the minus sign there are two cusps
 $(\pm\sqrt{2}+1,0)$.

The remaining SO(2,4) isometry generated by $P = 1_2$ and $Q = Q_2$
yields a configuration ${Y_0'}^2 - {Y_{-1}'}^2 = {Y_4'}^2 - {Y_1'}^2$ with
a slight sign difference from the starting solution (\ref{emy}).
In the Poincare coordinates it is given by
\begin{equation}
r^2 = y_0^2 - y_1^2 \pm \sqrt{2( y_0^2 + y_1^2 ) - 1 },
\label{fry}\end{equation}
whose surface ends on four lines $y_0 =y_1 \pm 1, y_0 = -y_1 \pm 1$
which meet at two cusps $(0, \pm1)$ for the plus sign and at two
cusps $(\pm1,0)$ for the minus sign.
The string surface (\ref{fry}) can be also confirmed to 
obey the string equation (\ref{ste}) in a similar way to the solution
(\ref{try}).

Now we consider the conformal SO(2,4) transformations that generate
a non-zero value of $y_2$. First we set $P = P_1$ and
\begin{equation}
Q = \left( \begin{array}{cccc} \cos\al & -\sin\al & 0 & 0 \\ 
0 & 0 & 0 & -1 \\  0 & 0 & 1 & 0 \\ \sin\al & \cos\al & 0 & 0 \end{array}
\right) \equiv Q_3(\al)  
\end{equation}
to have 
\begin{equation}
Y_0' Y_{-1}' = - \frac{\cos 2\al}{2}({Y_1'}^2 - {Y_2'}^2) +
\sin 2\al Y_1' Y_2', \hspace{1cm} Y_4' = 0,
\end{equation}
which give a surface 
\begin{eqnarray}
r &=& \sqrt{1 + y_0^2 - y_1^2 - y_2^2} \equiv \sqrt{A}, \nonumber \\
y_0 &=& - \frac{\cos 2\al}{2}(y_1^2 - y_2^2) + \sin 2\al y_1y_2.
\label{cs}\end{eqnarray}
At $\al = \pi/4$ this surface reduces to 
\begin{equation}
r = \sqrt{(1 - y_1^2)(1 - y_2^2)}, \hspace{1cm} y_0 = y_1y_2,
\label{fcu}\end{equation}
which show that the Wilson loop at the boundary is composed with 
four cusps and four lightlike lines and takes a square form for its 
projection on the $(y_1, y_2)$ plane \cite{AM}. 

We choose a static gauge
that $(y_1, y_2)$ are the worldsheet coordinates for the Euclidean
open string surface to express the Nambu-Goto action as
\begin{eqnarray}
S &=& \frac{R^2}{2\pi}\int  dy_1dy_2 \frac{\sqrt{D}}{r^2},
\nonumber \\
D &=& 1 + (\pa_ir)^2 - (\pa_iy_0)^2 - (\pa_1r \pa_2y_0 - 
\pa_2r\pa_1y_0)^2.
\label{dr}\end{eqnarray}
The equation of motion for $y_0$ is given by
\begin{equation}
\pa_i\left(\frac{\pa_iy_0}{r^2\sqrt{D}}\right) = \pa_1\left(
\frac{\pa_2rC}{r^2\sqrt{D}}\right) - \pa_2\left(\frac{\pa_1rC}
{r^2\sqrt{D}}\right), \hspace{1cm} C \equiv \pa_1r\pa_2y_0 - 
\pa_2r\pa_1y_0
\label{rdy}\end{equation}
and the equation of motion for $r$ takes a form
\begin{equation}
\frac{2}{r^3}\sqrt{D} = -\pa_i\left(\frac{\pa_ir}{r^2\sqrt{D}}\right) + 
\pa_1\left(\frac{\pa_2y_0C}{r^2\sqrt{D}}\right) - \pa_2\left(
\frac{\pa_1y_0C}{r^2\sqrt{D}}\right).
\label{cdr}\end{equation}
The insertion of the expression (\ref{cs}) into $C$ in (\ref{rdy})
and $D$ in (\ref{dr}) yields $C= -[2\cos2\al y_1y_2 + \sin2\al 
(y_1^2 - y_2^2)]/\sqrt{A}$ and $D = 1$. When the surface 
(\ref{cs}) is substituted
into the string equation (\ref{rdy}) its RHS can be so rewritten by
$(\pa_2r/r)\pa_1(C/r\sqrt{D}) -  (\pa_1r/r)\pa_2(C/r\sqrt{D})$ that it is
evaluated as $-2y_0(y_1^2 + y_2^2 - 2)/A^2$ which equals to the LHS
of (\ref{rdy}). The RHS of (\ref{cdr}) is expressed as sum
of two parts
\begin{eqnarray}
\frac{1}{A^{3/2}} [2- 3(y_1^2 + y_2^2)] + \frac{3}{A^{5/2}}[
\left( y_0( -\cos2\al y_1 + \sin2\al y_2 ) - y_1 \right)^2 \nonumber \\
+ \left( y_0( \cos2\al y_2 + \sin2\al y_1 ) - y_2 \right)^2 - 
(2\cos2\al y_1y_2 + \sin2\al (y_1^2 - y_2^2) )^2  ],
\end{eqnarray}
whose second part becomes proportional to $1/A^{3/2}$ and is summed up 
with the first part into $2/A^{3/2}$, which is the LHS of (\ref{cdr}).
For $\al = \pi/2$ or $(\al = 0)$ the string  surface (\ref{cs})
is given by
\begin{eqnarray}
y_0 &=& \frac{y_1^2 - y_2^2}{2}, \hspace{1cm} \left( y_0 = 
-\frac{y_1^2 - y_2^2}{2} \right), \nonumber \\
r &=& \sqrt{\left(1 + \frac{y_1 + y_2}{\sqrt{2}}\right)\left(1 - 
\frac{y_1 + y_2}{\sqrt{2}}\right)\left(1 + \frac{y_1 - y_2}{\sqrt{2}}
\right)\left(1 - \frac{y_1 - y_2}{\sqrt{2}}\right) },
\end{eqnarray}
from which we see that the Wilson loop at the boundary $r = 0$ has
a square-form projection on the $(y_1,y_2)$ plane characterized by the 
four lines, $y_2 = \pm y_1 + \sqrt{2}, y_2 = \pm y_1 - \sqrt{2}$ and 
contains four cusps located at $(y_0,y_1,y_2) = 
(1,\pm\sqrt{2},0), (-1,0,\pm\sqrt{2})$.
This square in the $(y_1,y_2)$ plane is produced by making a 
$\pi/4$-rotation of the square of the 4-cusp solution (\ref{fcu}).

Let us perform the SO(2,4) transformations specified by $P = 1_2,
Q = Q_3(\pi/4)$ and $P = -i\sigma_2, Q = Q_3(\pi/4)$ to derive two
string configurations
\begin{equation}
{Y_0'}^2 - {Y_{-1}'}^2 = -2Y_1' Y_2', \hspace{1cm} 
{Y_0'}^2 - {Y_{-1}'}^2 = 2Y_1' Y_2'
\end{equation}
with $Y_4'=0$, which are further represented by
\begin{eqnarray}
y_0 &=& \sqrt{1 - 2y_1y_2}, \hspace{1cm} r = \sqrt{2 - (y_1 + y_2)^2},
\label{bay} \\
y_0 &=& \sqrt{1 + 2y_1y_2}\equiv \sqrt{B}, \hspace{1cm} 
r = \sqrt{2 - (y_1 - y_2)^2} \equiv \sqrt{A}.
\label{ba}\end{eqnarray}
In order to see how the configuration (\ref{ba}), for instance, satisfies
the string equations we calculate $C$ and $\sqrt{D}$ to be compact
expressions $C = -(y_1^2 - y_2^2)/\sqrt{AB}, \sqrt{D} =1/\sqrt{B}$.
The RHS of (\ref{rdy}) is estimated as 
\begin{equation}
-\pa_1\frac{(y_1 - y_2)(y_1^2 - y_2^2)}{A^2} - \pa_2
\frac{(y_1 - y_2)(y_1^2 - y_2^2)}{A^2} = - \frac{2(y_1 - y_2)^2}{A^2},
\end{equation}
which agrees with the LHS of (\ref{rdy}). The RHS of (\ref{cdr})
can be separated into a $1/A^{3/2}$ part and a $1/A^{5/2}$ part as
\begin{equation}
\frac{1}{A^{3/2}\sqrt{B}}[2B- (y_1 - y_2)^2 - 2(y_1^2 + y_2^2)] +
\frac{1}{A^{5/2}\sqrt{B}}[ 6(y_1 - y_2)^2B - 3(y_1^2 - y_2^2)^2],
\label{ab}\end{equation}
whose second part becomes $3(y_1 - y_2)^2/A^{3/2}\sqrt{B}$ which makes
(\ref{ab}) equal to $2/A^{3/2}\sqrt{B}$, the LHS of (\ref{cdr}).

The projection of the  surface (\ref{ba}) at the boundary of
$AdS_5$ on the $(y_1,y_2)$ plane is composed of two separated parallel
lines, $y_2 = y_1 + \sqrt{2}$ on which $y_0 = |\sqrt{2}y_1 + 1|$
and $y_2 = y_1 - \sqrt{2}$ on which $y_0 = |\sqrt{2}y_1 - 1|$.
In the region defined by $A>0$, that is, the region between the 
two parallel lines, $B$  also takes a positive value.
From one cusp located at $(y_0,y_1,y_2) = (0,-1/\sqrt{2},1/\sqrt{2})$
two semi infinite lightlike lines emerge on a plane specified by
$y_2 = y_1 + \sqrt{2}$, while on a plane $y_2 = y_1 - \sqrt{2}$ two
semi infinite lightlike lines intersect transversly at the other 
cusp located at $(0,1/\sqrt{2},-1/\sqrt{2})$.
Thus the Wilson loop is composed of two separated parts each of which
is represented by two semi infinite lightlike lines with a cusp.
Therefore the solution (\ref{ba}) as well as (\ref{bay}) is regarded as
a two-cusp Wilson loop solution.

There remain two conformal transformations defined by $P = 1_2, 
Q = Q_3(\pi/2)$ and $P = 1_2, Q = Q_3(0)$, which produce the
following configurations
\begin{eqnarray}
{Y_0'}^2 - {Y_{-1}'}^2 &=& {Y_2'}^2 - {Y_1'}^2, \\
{Y_0'}^2 - {Y_{-1}'}^2 &=&  {Y_1'}^2 - {Y_2'}^2
\label{yyo}\end{eqnarray}
with $Y_4'=0$, which are respectively written by 
\begin{eqnarray}
y_0 &=& \sqrt{1 - y_1^2 + y_2^2}, \hspace{1cm} r = \sqrt{2 - 2y_1^2},
\label{yrb} \\
y_0 &=& \sqrt{1 + y_1^2 - y_2^2}\equiv \sqrt{B}, \hspace{1cm} r = 
\sqrt{2 - 2y_2^2} \equiv \sqrt{A}.
\label{yra}\end{eqnarray}
For the surface (\ref{yra}) $C$ and $\sqrt{D}$ are evaluated as
$C = 2y_1y_2/\sqrt{AB}, \sqrt{D} =1/\sqrt{B}$. In this case the 
demonstration of (\ref{rdy}) is simpler than the above cases due to 
$\pa_1r =0$. The eq. (\ref{cdr}) is also confirmed to hold in a 
way similar to (\ref{ab}). 

In ref. \cite{KRT} the solution (\ref{yyo}) has been analyzed in the
string sigma-model action in the conformal gauge and the leading
1-loop correction to it has been computed together with
the 1-loop correction to the starting 1-cusp solution (\ref{els}),
and further the 2-loop correction to the latter solution has been
studied \cite{RT}. In ref. \cite{KRT} the solution of (\ref{yyo}) was
presented by
\begin{eqnarray}
Y_0 &=& \frac{1}{\sqrt{2}} \cosh (\al \sigma + \beta \tau),
\hspace{1cm} Y_{-1} = \frac{1}{\sqrt{2}} \cosh (\al \tau - \beta\sigma),
\nonumber \\
Y_1 &=& \frac{1}{\sqrt{2}} \sinh (\al \sigma + \beta \tau),
\hspace{1cm} Y_2 = \frac{1}{\sqrt{2}} \sinh (\al \tau - \beta\sigma),
\hspace{1cm} Y_3 = Y_4 =0,
\label{coh}\end{eqnarray}
where $\al^2 + \beta^2 =2$, the Euclidean worldsheet coordinates 
$(\tau,\sigma)$ take values in the infinite interval.
The parametrization (\ref{coh}) is equivalently transformed in terms
of the Poincare coordinates into
\begin{eqnarray}
r &=& \frac{\sqrt{2}}{\cosh(\al \tau - \beta\sigma)}, \hspace{1cm}
y_0 = \frac{\cosh (\al\sigma + \beta \tau)}{\cosh(\al\tau - \beta\sigma)},
\nonumber \\
y_1 &=& \frac{\sinh(\al\sigma + \beta\tau)}{\cosh(\al\tau - \beta\sigma)},
\hspace{1cm} y_2 = \tanh (\al\tau - \beta\sigma),
\end{eqnarray}
which indeed satisfies the eq. (\ref{yra}).

The projection of the surface (\ref{yra}) at the boundary of $AdS_5$,
$r=0$ on the $(y_1,y_2)$ plane is composed of two separated parallel
lines, $y_2 = 1$ and $y_2 = -1$ on which $y_0$ is specified by the same
equation $y_0 = |y_1|$. From a different viewpoint the projection
of the surface (\ref{yra}) on the $(y_0,y_1)$ plane at a fixed
value of $y_2$ in $-1 < y_2 < 1$ is described by a hyperbolic curve 
$y_0 = \sqrt{y_1^2 +(1 -y_2^2)}$, while that projection at the
boundary value $y_2 =1$ or $y_2 =-1$ becomes two semi infinite
lightlike lines intersecting transversly at a cusp located
at $(y_0,y_1,y_2) = (0,0,1)$ or $(0,0,-1)$.
Thus the solution (\ref{yra}) as well as (\ref{yrb}) has two cusps
in the same way as (\ref{ba}) and (\ref{bay}).

\section{Conformal boosts of the 4-cusp Wilson loop solution}

We turn to the conformal boost transformations of the 4-cusp solution
$Y_0Y_{-1} = Y_1Y_2, Y_3=Y_4=0$ (\ref{fcu}) and see whether the 
transformed configurations are solutions of the string equations for the
Nambu-Goto action. The boost in the (-1,4) plane given by
\begin{equation}
Y_4 = \gamma(Y_4' - vY_{-1}'), \hspace{1cm} 
Y_{-1} = \gamma(Y_{-1}' - vY_4')
\label{vb}\end{equation}
with $\gamma = 1/\sqrt{1-v^2}$ is performed to yield $Y_4' - vY_{-1}'=0$
and $\gamma Y_0'(Y_{-1}' - vY_4')=Y_1'Y_2'$, which are represented
in terms of the Poincare coordinates as
\begin{equation}
y_0' = \gamma(1+v)y_1'y_2', \hspace{1cm} r' = \sqrt{
\frac{1-v}{1+v} + {y_0'}^2 - {y_1'}^2 - {y_2'}^2 } \equiv \sqrt{A}.
\label{ypr}\end{equation}
Alternatively the boost (\ref{vb}) can be described by
\begin{equation}
r' = \sqrt{\frac{1-v}{1+v}}r, \hspace{1cm} y_1' = \sqrt{\frac{1-v}{1+v}}
y_1, \hspace{1cm} y_2' = \sqrt{\frac{1-v}{1+v}}y_2
\label{rp}\end{equation}
owing to the relation $1/r' = Y_{-1}' + Y_4' = \gamma (1+v)/r$.
Below the prime will be omitted. For the eq. (\ref{rdy}) $C$ is given by
$C=-\gamma(1+v)(y_1^2 -y_2^2)/\sqrt{A}$, while $D$ takes a simple value
$D=1$. The RHS of (\ref{rdy}) is computed by
\begin{eqnarray}
-\frac{2\gamma(1+v)}{A^3}\Biggl[A\left( \gamma^2(1+v)^2y_1y_2
(y_1^2 + y_2^2) -   2y_1y_2 \right) \nonumber \\
+ (y_1^2 - y_2^2)\left( -(\gamma^2(1+v)^2y_1^2 -1 )y_2\pa_1A
+ (\gamma^2(1+v)^2y_2^2 -1 )y_1\pa_2A \right) \Biggr],
\end{eqnarray}
whose $1/A^3$ part vanishes owing to the symmetric form of $A$
for the exchange of $y_1$ and $y_2$. The $1/A^2$ part becomes equal to
the LHS of (\ref{rdy}). The RHS of the other string equation (\ref{cdr})
is calculated by
\begin{equation}
\frac{1}{A^{3/2}}[2- 3\gamma^2(1+v)^2(y_1^2 + y_2^2)] + 
\frac{3(y_1^2 + y_2^2)}{A^{5/2}}[\gamma^4(1+v)^4y_1^2y_2^2
+ 1 - \gamma^2(1+v)^2(y_1^2 + y_2^2)],
\label{var}\end{equation}
whose second $1/A^{5/2}$ part is so simplified into 
$3(y_1^2 + y_2^2)(1+v)/(A^{3/2}(1-v))$ that (\ref{var}) becomes
$2/A^{3/2}$ which is just the LHS of (\ref{cdr}).

Since $r'$ of (\ref{ypr})  is represented with the prime by
\begin{equation}
r' = \sqrt{\frac{1 + v}{1 - v}} \sqrt{\left( \frac{1-v}{1+v} - {y_1'}^2
\right)\left(  \frac{1-v}{1+v} - {y_2'}^2\right) },
\label{rvy}\end{equation} 
the projection of the string surface on the $(y_1', y_2')$ plane at
the $AdS_5$ boundary is the square with  side length $2\sqrt{(1-v)/(1+v)}
\equiv 2a$, which is compared to the square with  side length 2 for the
starting basic solution (\ref{fcu}). Following the IR dimensional 
regularization of ref. \cite{AM} we define the following 
regularized Nambu-Goto action
\begin{equation}
S = \frac{\sqrt{\la_d c_d}}{2\pi} \int dy_1' dy_2' 
\frac{\sqrt{D}}{{r'}^{2+ \ep}}
\label{acd}\end{equation}
with $d = 4-2\ep, c_d= 2^{4\ep}\pi^{3\ep}\Gamma(2+\ep),
\la_d = \la\mu^{2\ep}/(4\pi e^{-\gamma})^{\ep}, \gamma = -\Gamma'(1)$,
where $\la_d$ is described by the IR cutoff scale $\mu$ and
the dimensionless four dimensional coupling $\la$ to match the
field theory side. Substituting the solution (\ref{rvy}) into the action
(\ref{acd}) and making a variable transformation (\ref{rp})  to
carry out an integral over the inside of the square we have
\begin{equation}
-iS = B_{\ep} \int_{-1}^1 dy_1dy_2 \frac{1}{[(1-y_1^2)(1-y_2^2)]^{1 + 
\frac{\ep}{2} }} =  B_{\ep} \frac{\pi\Gamma\left(-\frac{\ep}{2}\right)^2}
{\Gamma\left(\frac{1-\ep}{2}\right)^2 }
\label{sep}\end{equation}
with $B_{\ep} = \sqrt{\la_d c_d}/2\pi a^{\ep}$, where $i$ is due to
the Euclidean worldsheet and a double pole appears when $\ep < 0 
\rightarrow 0$. From the structure of the 
action (\ref{dr}) if $r$ and $y_{\mu}$
are solutions of the string equations, then the rescaling configuration
specified by $ar$ and $ay_{\mu}$ with an arbitrary constant $a$ also 
satisfies the string equations. The solution (\ref{rp}) exhibits a 
rescaling solution.

Let us perform the following boost in the (0,4) plane for the
4-cusp solution (\ref{fcu})
\begin{equation}
Y_4 = \gamma(Y_4' - vY_0'), \hspace{1cm} Y_0 = \gamma(Y_0' - vY_4').
\label{ybt}\end{equation} 
We obtain a string surface described by
\begin{equation}
r' = \sqrt{ 1 - \frac{2b}{\gamma}y_0' + {y_0'}^2 - {y_1'}^2 - {y_2'}^2 }
\equiv \sqrt{A}, \hspace{1cm} b = \gamma v
\label{gab}\end{equation}
and $v{y_0'}^2 - y_0' + \gamma y_1'y_2' =0$ for which we choose
\begin{equation}
y_0' = \frac{\gamma}{2b} ( 1- \sqrt{1- 4by_1'y_2'} )
\equiv \frac{\gamma}{2b} ( 1- \sqrt{B} ),
\label{bga}\end{equation}
which reduces to the starting solution (\ref{fcu}) in the limit $v 
\rightarrow 0$. We consider the $b < 1$ case. 
Below the prime will be suppressed. For the eq. 
(\ref{rdy}) $C$ is evaluated as $C = -\gamma(y_1^2 - y_2^2)/\sqrt{AB}$
from the involved expressions (\ref{gab}) and (\ref{bga}), while
$D$ in (\ref{dr}) has many complicated terms but can be cast into
a simple form $\sqrt{D} = 1/\sqrt{B}$ through $\gamma^2 = 1+ b^2$.
The substitution of these 
expressions into the RHS of (\ref{rdy}) leads to
\begin{eqnarray}
&\frac{\gamma}{A^2}\left[ 2b\left( 1- \frac{\gamma^2}{2b^2} \right)
\frac{y_1^2 + y_2^2}{\sqrt{B}} + 4y_1y_2 + \frac{\gamma^2}{b}
(y_1^2 + y_2^2) \right]&  \\
&+ \frac{2\gamma}{A^3}(y_1^2 - y_2^2)
\left[ \left( \left(1- \frac{\gamma^2}{2b^2} \right)
\frac{by_2}{\sqrt{B}} + y_1 + \frac{\gamma^2}{2b}y_2 \right)\pa_2A
- \left( \left(1- \frac{\gamma^2}{2b^2} \right)
\frac{by_1}{\sqrt{B}} + y_2 + \frac{\gamma^2}{2b}y_1 \right)\pa_1A  
\right],& \nonumber
\end{eqnarray}
whose second $1/A^3$ part vanishes owing to the symmetric expression of
$A$ for the exchange of $y_1$ and $y_2$ and the remaining first part
becomes coincident with the LHS of (\ref{rdy}).
In order to prove the eq. (\ref{cdr}) we devote ourselves to the
$1/A^{5/2}$ part of its RHS
\begin{eqnarray}
\frac{3}{A^{5/2}\sqrt{B}}\Biggl[ \left( 1- \frac{\gamma^2}{2b^2} \right)^2
b^2(y_1^2 + y_2^2) +2b\sqrt{B}\left( 1- \frac{\gamma^2}{2b^2} \right)
\left(2y_1y_2 + \frac{\gamma^2}{2b}(y_1^2 + y_2^2) \right) 
\nonumber \\
+ (1 - 4by_1y_2 )\left( \frac{2\gamma^2}{b}y_1y_2 + \left( 1 + 
\frac{\gamma^4}{4b^2} \right)(y_1^2 + y_2^2) \right)
- \gamma^2(y_1^2 - y_2^2)^2 \Biggr],
\end{eqnarray}
which turns out to be $6b(2y_1y_2 + \gamma^2(y_1^2 + y_2^2)/2b )/A^{3/2}
\sqrt{B}$, which further cancels against the remaining $1/A^{3/2}$ part
to leave $2/A^{3/2}\sqrt{B}$, that is, the LHS of (\ref{cdr}).

The insertion of (\ref{bga}) into (\ref{gab}) with $r'=0$ leads to
the projection of the Wilson loop on the $(y_1',y_2')$ plane, which is
expressed as
\begin{equation} 
({y_1'}^2 + {y_2'}^2)^2 - \frac{\gamma^2}{b^2}({y_1'}^2 + {y_2'}^2)
(1 - 2by_1'y_2' ) + \frac{\gamma^4}{b^2}{y_1'}^2{y_2'}^2
- \frac{4}{b}y_1'y_2' + \frac{1}{b^2} = 0.
\end{equation}
This equation can be factorized into
\begin{equation}
\left[ \left( y_2' + \frac{y_1'}{b} \right)^2 - \frac{1}{b^2} \right]
\left[ ( y_2' + by_1')^2 - 1 \right] = 0,
\label{fac}\end{equation}
which gives four lines $y_2' + y_1'/b = \pm 1/b, y_2' + by_1' = \pm1$
that form a rhombus on the $(y_1',y_2')$ plane.

The boost (\ref{ybt}) gives $1/r' = Y_{-1}' + Y_4' = Y_{-1} + 
\gamma(Y_4 + vY_0)$ and $y_0'/r' = \gamma(Y_0 + vY_4)$, which
become $1/r' = (1 + by_0)/r$ and $y_0' = \gamma y_0r'/r$ through the 
starting solution (\ref{fcu}). Thus the boost is represented in terms of
the Poincare coordinates as
\begin{equation}
y_0' = \frac{\gamma y_1y_2}{1 + by_1y_2}, \hspace{1cm} y_1' = 
\frac{y_1}{1 + by_1y_2}, \hspace{1cm} y_2' = \frac{y_2}{1 + by_1y_2}.
\label{gap}\end{equation}
The eq. (\ref{bga}) is changed into the first equation in (\ref{gap}) when
the second and third equations in (\ref{gap})
are substituted into (\ref{bga}), while the eq. (\ref{gab}) becomes
\begin{equation}
r' = \frac{\sqrt{(1-y_1^2)(1-y_2^2)} }{1 + by_1y_2},
\label{rsq}\end{equation}
which vanishes at $y_1 = \pm 1$ and $y_2 = \pm1$. The four cusps of the
Wilson loop are determined from (\ref{gap}) with $y_1 = \pm 1, 
y_2 = \pm1$ to be located in the coordinates $(y_0',y_1',y_2')$ as
\begin{eqnarray}
\left( -\frac{\sqrt{1 + b^2}}{1- b}, -\frac{1}{1- b}, \frac{1}{1- b} 
\right), \hspace{1cm}  \left( \frac{\sqrt{1 + b^2}}{1+ b}, \frac{1}{1+ b},
\frac{1}{1 + b} \right), \nonumber \\
\left( \frac{\sqrt{1 + b^2}}{1 + b}, -\frac{1}{1 + b}, -\frac{1}{1 + b} 
\right), \hspace{1cm}  \left(-\frac{\sqrt{1 + b^2}}{1- b}, \frac{1}{1- b},
-\frac{1}{1 - b} \right).
\label{pos}\end{eqnarray}
We again see that the projection of the Wilson loop on the $(y_1',y_2')$
plane is a rhombus. Alternatively the cusp positions (\ref{pos}) are
determined from the intersections of the four lines defined by 
(\ref{fac}). The four lightlike segments between the adjacent cusps
characterize the four massless gluon momenta so that the parameter $b$ is
related with the ratio of the Mandelstam variables $s, t$ as
$s/t = (1 + b)^2/(1 - b)^2$. The classical Nambu-Goto action (\ref{acd})
evaluated on this 4-cusp Wilson loop solution is represented through
(\ref{gap}) and (\ref{rsq}) as
\begin{eqnarray}
-iS &=& \frac{\sqrt{\la_d c_d}}{2\pi} \int_{-1}^1 dy_1dy_2 \frac{1 - 
by_1y_2}{(1 + by_1y_2)^3} \frac{1}{r'^{2+\ep} \sqrt{B(y_i')}} \nonumber \\
 &=& \frac{\sqrt{\la_d c_d}}{2\pi} \int_{-1}^1 dy_1dy_2 
\frac{(1 + by_1y_2)^{\ep}}{[(1 - y_1^2)(1 - y_2^2)]^{1+\frac{\ep}{2}} }.
\end{eqnarray}
By expanding the integrand as a power series on $b$  we evaluate the
integral over $y_1$ and $y_2$ as
\begin{equation}
\frac{\sqrt{\la_d c_d}}{2\pi} \frac{\pi \Gamma\left(
-\frac{\ep}{2}\right)^2}{\Gamma\left(\frac{1-\ep}{2}\right)^2} F\left(
\frac{1}{2}, -\frac{\ep}{2}, \frac{1-\ep}{2}; b^2 \right).
\label{fgp}\end{equation}

Here in view of (\ref{gab}) and (\ref{bga}) we present a string 
configuration using an arbitrary constant $a$
\begin{eqnarray}
r' &=& \sqrt{ a^2 - \frac{2ab}{\gamma}y_0' + {y_0'}^2 - {y_1'}^2 - 
{y_2'}^2 } \equiv \sqrt{A}, \hspace{1cm} b = \gamma v,
\label{ary} \\
y_0' &=& \frac{\gamma}{2b} ( a - \sqrt{a^2- 4by_1'y_2'} )
\equiv \frac{\gamma}{2b} ( a- \sqrt{B} ),
\label{abb}\end{eqnarray}
whose $v \rightarrow 0$ limit reduces to (\ref{ypr}) with $a$ replaced by
$\sqrt{(1-v)/(1+v)}$. This string surface is confirmed to satisfy
the string equations with $C=-\gamma(y_1^2 - y_2^2)/\sqrt{AB}$ and
$\sqrt{D} = a/\sqrt{B}$. This solution is just a rescaling solution
for (\ref{gap}) and (\ref{rsq}) as expressed by
\begin{eqnarray}
y_0' &=& \frac{a\gamma y_1y_2}{1 + by_1y_2}, \hspace{1cm} y_1' = 
\frac{ay_1}{1 + by_1y_2}, \hspace{1cm} y_2' = \frac{ay_2}{1 + by_1y_2},
\label{ayb} \\
r' &=& \frac{a\sqrt{(1-y_1^2)(1-y_2^2)} }{1 + by_1y_2}.
\label{ray}\end{eqnarray}
The insertion of (\ref{ayb}) into (\ref{ary}) leads to (\ref{ray}) and
the eq. (\ref{abb}) becomes the first equation in (\ref{ayb}) when
the second and third equations are substituted.
The classical action of this rescaling solution is similarly
evaluated as
\begin{equation}
-iS = \frac{\sqrt{\la_d c_d}}{2\pi} \int_{-1}^1 dy_1dy_2 
\frac{a^2(1 - by_1y_2)}{(1 + by_1y_2)^3} 
\frac{a}{r'^{2+\ep} \sqrt{B(y_i')}},
\end{equation}
which gives (\ref{fgp}) with the $\sqrt{\la_d c_d}/2\pi$ factor replaced
by $B_{\ep}=\sqrt{\la_d c_d}/2\pi a^{\ep}$, whose $\ep$ expansion leads to
the exponential form of the planar 4-gluon scattering amplitude at strong
coupling of \cite{AM}. In ref. \cite{AM} for the 4-cusp Wilson loop
solution the string sigma-model action was used, while 
based on the Nambu-Goto action we have demonstrated that the 4-cusp Wilson
loop configuration indeed solves the string equation and evaluated
the classical action on this 4-cusp solution.

There is another boost in the (-1,1) plane specified by
\begin{equation}
Y_1 = \gamma(Y_1' - vY_{-1}'), \hspace{1cm} Y_{-1} = 
\gamma(Y_{-1}' - vY_1'),
\label{gvy}\end{equation} 
which produces a string surface
\begin{equation}
y_0' = \frac{y_2'(y_1' - v)}{1 - vy_1'} \equiv \frac{y_2'(y_1' - v)}{B},
\hspace{1cm} r' = \sqrt{ 1 + {y_0'}^2 - {y_1'}^2 - {y_2'}^2 }
\equiv \sqrt{A}.
\label{yby}\end{equation}
This configuration is not symmetric under the interchange of $y_1'$ and
$y_2'$. Although $C$ is obtained by an involved form $C = [-y_1(y_1-v)/B
+ (1-v^2)y_2^2/B^2]/\sqrt{A}$ here with unprimed expressions,
$\sqrt{D}$ is calculated to be a simple form $\sqrt{D}=\sqrt{1-v^2}/B$.
The string equation (\ref{rdy}) is proved as follows.
The RHS of (\ref{rdy}) is evaluated as
\begin{eqnarray}
&\frac{y_2}{A^2} \left[ \sqrt{1-v^2}\left( 2y_1X - (1-y_1^2)\pa_1X \right)
- \frac{2B^2}{\sqrt{1-v^2}} \left( \frac{(1-v^2)(y_1 - v)}{B^3}X
+ \left( \frac{(1-v^2)y_2^2(y_1 - v)}{B^3} - y_1 \right)
\frac{1-v^2}{B^3}  \right) \right]& \nonumber \\
&+\frac{2X}{A^3} \left[ \sqrt{1-v^2}y_2(1-y_1^2)\pa_1A +
\frac{B^2}{\sqrt{1-v^2}}\left( \frac{(1-v^2)y_2^2(y_1 - v)}{B^3} - y_1 
\right)\pa_2A \right], &
\end{eqnarray}
where $X = -y_1(y_1-v)/B^2 + (1-v^2)y_2^2/B^3$ and the $1/A^3$
part vanishes through $A = (1-y_1^2)(1 - y_2^2(1-v^2)/B^2)$
in spite of the asymmetric expression. The remaining $1/A^2$ part agrees
with the LHS of (\ref{rdy}). For the string equation (\ref{cdr}),
the $1/A^{5/2}$ terms of the RHS are gathered together into a sum of
terms with odd powers of $1/B$
\begin{equation}
\frac{3\sqrt{1-v^2}(1-y_1^2)}{A^{5/2}B} \left[ y_1^2 +
\frac{(1-v^2)y_2^2(1 - y_1^2)}{B^2} - \frac{(1-v^2)^2y_2^4}{B^4} \right],
\end{equation}
which turns out to be a $1/A^{3/2}$ term as 
$3\sqrt{1-v^2}( y_1^2 + (1-v^2)y_2^2/B^2)/BA^{3/2}$. It combines with
the remaining $1/A^{3/2}$ terms to be equated with 
$2\sqrt{1-v^2}/BA^{3/2}$, that is, the LHS of (\ref{cdr}), where
the $1/B^3$ terms are canceled out and the $1/B^2$ terms can be
changed into a $1/B$ term. 

The string solution (\ref{yby}) can be expressed as 
\begin{equation}
r' = \frac{1}{1 - vy_1'}\sqrt{ (1- {y_1'}^2) \left( (1-vy_1')^2
- (1-v^2){y_2'}^2 \right) }
\end{equation}
so that the projection of the Wilson loop on the $(y_1', y_2')$ plane
is a trapezium formed by the four lines $y_1' = \pm1$ and 
$y_2' = \pm(vy_1' - 1)/\sqrt{1-v^2}$. In $(y_0',y_1',y_2')$ the four
cusps are located at
\begin{eqnarray}
\left(-\sqrt{ \frac{1+v}{1-v}}, -1, \sqrt{\frac{1+v}{1-v}} \right), 
\hspace{1cm}  \left( \sqrt{ \frac{1-v}{1+v}}, 1, \sqrt{\frac{1-v}{1+v}}
 \right), \nonumber \\
\left(\sqrt{ \frac{1+v}{1-v}}, -1, -\sqrt{\frac{1+v}{1-v}} \right), 
\hspace{1cm}  \left( -\sqrt{ \frac{1-v}{1+v}}, 1, -\sqrt{\frac{1-v}{1+v}}
 \right),
\label{fov}\end{eqnarray}
which imply that the Wilson loop consists of the four lightlike segments.
The boost (\ref{gvy}) is alternatively expressed as
\begin{equation}
y_0' = \frac{y_1y_2}{\gamma(1 + vy_1)}, \hspace{1cm} y_1' = \frac{v + y_1}
{1 + vy_1}, \hspace{1cm} y_2' = \frac{y_2}{\gamma(1 + vy_1)}, 
\end{equation}
which lead to  (\ref{fov}) again and are substituted into the
second eq. of (\ref{yby}) to be
\begin{equation}
r' = \frac{\sqrt{(1-y_1^2)(1-y_2^2)}}{\gamma(1 + vy_1)}.
\end{equation}
Then the action (\ref{acd}) can be evaluated on this classical solution as
\begin{eqnarray}
-iS &=& \frac{\sqrt{\la_d c_d}}{2\pi} \int_{-1}^1 dy_1dy_2 \frac{1 - 
v^2}{\gamma(1 + vy_1)^3} \frac{\sqrt{1-v^2}}{r'^{2+\ep} B(y_1')} 
\nonumber \\
 &=& \frac{\sqrt{\la_d c_d}}{2\pi} \int_{-1}^1 dy_1dy_2 
\frac{(\sqrt{1 + b^2} + by_1)^{\ep}}
{[(1 - y_1^2)(1 - y_2^2)]^{1+\frac{\ep}{2}} }.
\end{eqnarray}
The integral is calculated by expanding the integrand as a power
series on $b$ as
\begin{equation}
\sqrt{\pi} \frac{\Gamma\left(\frac{1}{2}\right)\Gamma\left(
-\frac{\ep}{2}\right)}{\Gamma\left(\frac{1- \ep}{2}\right)^2}
(1 + b^2 )^{\ep/2} \sum_{n=0}^{\infty}\left( \frac{b^2}{1 + b^2}
\right)^n \frac{\Gamma\left(n- \frac{\ep}{2} \right) }{n!},
\end{equation}
whose summation factor is evaluated as $\Gamma(-\ep/2)
(1- b^2/(1+b^2))^{\ep/2}$ through $(1-z)^{\al} = F(-\al,\beta,\beta,z)$
so that we obtain $\pi\Gamma(-\ep/2)^2/\Gamma((1-\ep)/2)^2$.
This result is compared with (\ref{sep}) with $a=1$ or (\ref{fgp})
with $b=0$, the Mandelstam variables $s=t$, which is consistent
with the observation that the locations (\ref{fov}) of the four
cusps indicate $s=t$. 

Here we discuss the remaining boost in the (0,1) plane defined by
\begin{equation}
Y_1 = \gamma(Y_1' - vY_0'), \hspace{1cm} Y_0 = \gamma(Y_0' - vY_1'),
\end{equation} 
which yields a string configuration
\begin{equation}
y_0' = \frac{y_1'(y_2' + v)}{1 + vy_2'},
\hspace{1cm} r' = \sqrt{ 1 + {y_0'}^2 - {y_1'}^2 - {y_2'}^2 }.
\label{zrv}\end{equation}
For (\ref{zrv}) the projection of the Wilson loop on the $(y_1',y_2')$
plane shows a trapezium in the same way as that for (\ref{yby}).
Although this configuration is transformed into (\ref{yby}) under 
the interchanges $y_1'  \leftrightarrow y_2'$ and 
$v \leftrightarrow -v$, we can show that it satisfies the string 
equations in the same way as the solution (\ref{yby}).

\section{Conclusions}

Starting with the elementary 1-cusp Wilson loop solution of \cite{MK},
we have constructed various kinds of string configurations by performing
the conformal SO(2,4) transformations in the embedding coordinates
of $AdS_5$ and then rewriting the results back in the Poincare 
coordinates. Analyzing the obtained string surfaces to see where they
end on at the $AdS_5$ boundary, we have read off the shapes of the
Wilson loops. In order to see whether the conformal transformed string
configurations are extrema of the worldsheet area we have demonstrated
that they indeed solve the involved string equations of motion for the 
Nambu-Goto string action. In these demonstrations the string Lagrangian
$\sqrt{D}/r^2$ does not take a constant value in our worldsheet
coordinates but it is important that $\sqrt{D}$ takes a simple 
manageable expression.

We have made two types of SO(2)$\times$SO(4) transformations that are
characterized by two kinds of SO(4) rotations such that one does not
change $Y_2$ and the other interchanges $Y_2$ and $Y_4$.
We have observed that the former type of transformaions generate
a variety of 2-cusp Wilson loop solutions, while the latter type of them
produce not only the 4-cusp Wilson loop solutions but also the
2-cusp Wilson loop solutions. The projection of the latter
2-cusp Wilson loop surfaces on the $(y_1,y_2)$ 
plane is two separated parallel lines,
which is compared with the square-form projection of the 4-cusp
Wilson loop surface. Applying the boost SO(2,4) transformations to the
basic 4-cusp solution with the square-form projection we have constructed
three kinds of 4-cusp solutions whose projections are the 
rescaled square, the rhombus and the trapezium.

By combining the conformal boost in the (0,4) plane and the rescaling
we have derived a 4-cusp Wilson loop surface whose
projection is a rescaled rhombus. Based on the Nambu-Goto
action in the IR dimensional regularization we have obtained the classical
Euclidean action evaluated on this 4-cusp solution and reproduced
the same exponential expression of the 4-gluon amplitude as 
derived in \cite{AM} from the string sigma-model action.

\end{document}